# A dynamic modeling tool for estimating healthcare demand from the COVID19 epidemic and evaluating population-wide interventions

**Authors:** Gabriel Rainisch[1], Eduardo A. Undurraga[2,3], Gerardo Chowell[1]

**Author Affiliations:**

[1] Department of Population Health Sciences, School of Public Health, Georgia State University, Atlanta, GA, USA),

[2] Escuela de Gobierno, Pontificia Universidad Católica de Chile, RM, Chile

[3] Millennium Initiative for Collaborative Research in Bacterial Resistance (MICROB-R), RM, Chile

**ABSTRACT**

**Objectives.** Public health officials need tools to assist with anticipating the healthcare resources required to confront the SARS-COV-2 pandemic. We built a modeling tool to aid practicing public health officials with estimating healthcare demand from the pandemic in their jurisdictions and to evaluate the potential impacts of population-wide social-distancing interventions.

**Methods.** The tool uses a SEIR compartmental model to project the pandemic's local spread. Users input case counts, healthcare resources, and select intervention strategies to evaluate. Outputs include the number of infections and deaths with and without intervention, and the demand for hospital and critical care beds and ventilators relative to existing capacity. We illustrate the tool using data from three regions of Chile.

**Results.** Our scenarios indicate a surge in COVID-19 patients could overwhelm Chilean hospitals by June, peaking in July or August at 6 to 50 times the current supply of beds and ventilators. A lockdown strategy or combination of case isolation, home quarantine, social distancing of individuals >70 years, and telework interventions may keep treatment demand below capacity.

**Conclusions.** Aggressive interventions can avert substantial morbidity and mortality from COVID-19. Our tool permits rapid evaluation of locally-applicable policy scenarios and updating of results as new data become available.

# 1. Introduction

On December 31, 2019, the regional office of the World Health Organization (WHO) was notified of a cluster of pneumonia cases of unknown origin associated with a market in Wuhan, China (Zhu et al., 2020). A novel coronavirus (SARS-COV-2) was identified as the cause of the infections (Zhu et al., 2020) and has since spread worldwide. As of May 7, 2020, more than 3.6 million cases of COVID-19 (illness caused by SARS-COV-2) have been reported in 184 countries and territories, including ~250,000 deaths (Dong et al., 2020, World Health Organization, 2020). The pandemic has overwhelmed both national and local healthcare capacity in several countries (Ferguson et al., 2020, Kissler et al., 2020), and is projected to do so in many others. Low- and middle-income countries are particularly vulnerable (Walker et al., 2020), since financial and logistical challenges may hinder their ability to augment treatment capacity. As such, many countries have resorted to societal-wide social distancing interventions in the hopes of reducing morbidity and delaying the demand for healthcare resources, in order to gain time to increase treatment capacity.

Numerous modeling efforts have forecasted the spread of the outbreak and examined the potential benefits of social-distancing interventions (Ferguson et al., 2020, Flaxman et al., 2020, Kissler et al., 2020, Walker et al., 2020). While informative, these efforts have been limited to specific nations and snapshots in time and public health officials are reliant on the authors for updated estimates as the pandemic evolves. Other internet-based tools offer public health users the ability to generate estimates on their own, but these are limited in their practical utility because their assumptions and desired results may not match the specific needs of jurisdictions and public health decision makers, or they require coding knowledge to access or modify the calculations (Healthcare, 2020, Henderson, 2020). These considerations are more critical in low and middle-income countries, which may not have the resources to complete or modify such analyses on their own.

Therefore, we developed a modeling tool for use by practicing public health officials to estimate the future impact of the CoVID-19 outbreak on the demand for healthcare resources in their jurisdictions and for examining the costs and benefits of various intervention strategies. Once downloaded, the model can be used without an internet connection, to assist public health officials with choosing locally applicable intervention strategies and by how much to increasing hospital treatment capacity. For illustration, we apply the model to Chile, a Southern Hemisphere country where the virus is generating local transmission, and compare various interventions options in the three most affected regions of the country.

## 2. Methods

*2.1. Tool Overview*

We created a spreadsheet-based tool (Supplementary Material S1) that uses a Susceptible-Latent-Infectious-Recovered (SEIR) Compartmental Model to project the future impact of a COVID-19 epidemic among any population of interest. The model requires information that is typically available for public health officials, including the number of cases in their jurisdiction, the size and demographics of their at-risk population, healthcare capacity, expectations for healthcare use, and choices of societal-wide social-distancing mitigation strategies users wish to evaluate. Model outputs reflect the potential demand on the healthcare system due to severely ill individuals with and without user-specified mitigation strategies, as well as deaths averted through treatment and excess deaths due to healthcare demand exceeding capacity. The demand for healthcare resources is measured as the estimated number of CoVID-19 patients requiring critical-care or Intensive Care Unit (ICU) beds, hospital beds (non-ICU), and mechanical ventilators over the course of the outbreak and the maximum occupancy at the outbreak's peak. The tool offers users the ability to evaluate various intervention strategies currently under consideration and in use worldwide (Ferguson et al., 2020, Kissler et al., 2020, Willem et al., 2020). These interventions comprise five mitigation-type interventions which focus on slowing epidemic spread and reducing its burden on the healthcare system, and one suppression-type strategy, which employs aggressive interventions aimed at reversing epidemic growth (Table 1).

[Table 1 about here]

Users can readily update all input values as new data become available or reflect a jurisdiction's specific epidemiologic profile of disease and policy considerations. All calculations can be readily modified by users (although no modifications are necessary for tool use).

*2.2. Calculations*

*2.2.1. Transmission with and without intervention*

Our SEIR model tracks the number of individuals transitioning between disease states every day of the outbreak. The initial number of susceptible individuals is set as the population minus the cumulative number infected since the outbreak's start. Transmission occurs through contacts between susceptible and infectious individuals and we assume an equal probability any one person has contact with another ("homogenous mixing"). We also assume transmission chains generated by infected travelers entering the population are minimal compared to existing transmission in the community. As a result, the number of new infections each day is the product of the proportion of the population that is susceptible, the number of infectious persons on a given day, and the average number of new infections each infected person causes over the span of their illness (the reproduction number; hereafter, "R") divided by the duration (in days) of the average infectious period. Infectiousness is assumed to occur five days after infection (Lauer et al., 2020) and lasts 11 days (You et al., 2020). Upon recovery from infection, individuals are assumed immune to re-infection during the timespan modeled (through December 2020). In the absence of intervention R is 2.0 (low estimate) and 2.8 (high estimate), approximately spanning the middle 50% of the gamma distribution of R (95% intervals: 1.4-3.8) estimated from the early growth-rate of the epidemic in Wuhan (Li et al., 2020, Riou and Althaus, 2020). To account for uncertainty in R, all results are depicted as a range based on these low and high estimates for R. During time periods where interventions are applied, we reduce the low and high estimates of R by the values in Table 1. Upon mitigation concluding, R returns to pre-mitigation levels to illustrate the potential consequences of shorter duration interventions. However, advanced users can alter the tool so that when one mitigation strategy concludes, another begins. Finally, we do not account for any vaccine as it is only likely to be available beyond the modeled time frame (Li and De Clercq, 2020, Nature, 2020). All equations governing dynamics of the system are provided in the supplementary material.

*2.2.2. Hospitalizations and ICU admissions with and without intervention*
In our model, all symptomatic persons with an illness severe enough to warrant hospitalization will seek healthcare and the risk for hospitalization is age-dependent (Table 2) (Verity et al., 2020). Similarly, the percentages of individuals admitted to the hospital requiring ICU care and fatality are also age-dependent (Table 2), while the likelihood of patients admitted to the ICU who require mechanical ventilation is assumed the same (63.2%) for all ages (ICNARC, 2020).

Based on observations for COVID19, we assume individuals seeking hospital care do so 11 days after infection (five days incubation + six days of symptoms) (Chen et al., 2020, Li et al., 2020, Linton et al., 2020, Zhang Guqin et al., 2020). We calculate hospital (non-ICU) bed occupancy based on a ten day length of stay for patients treated entirely in non-critical hospital wards (Deng et al., 2020, Wang Dawei et al., 2020) and ICU bed occupancy based on a ten day length of stay when critical care is required (Verity et al., 2020, Zhang Guqin et al., 2020). We assume a four day lag from hospital admission to ICU admission (Wang Dawei et al., 2020, Zhang Guqin et al., 2020). When mechanical ventilation is required, we assume the duration of use is nine days, based on expert clinical opinion that ventilation is necessary for the duration of ICU stays other than two days (one-day lag post ICU admission to initiate ventilator use plus 1 day in the ICU post-use) and another day required for ventilator cleaning/re-equipping.

[Table 2 about here]

To estimate the impact of interventions on hospital resource requirements we calculate two measures for each of the three resources tracked in the model: 1) the reduction in peak occupancy between the projected outbreak without intervention and when interventions are employed, and 2) the number of weeks peak occupancy is delayed due to employed interventions.

*2.2.3. Deaths with and without intervention*

We assume all deaths occur in the hospital unless treatment capacity is overwhelmed, and that it takes the same time for an individual to recover and die, despite some preliminary evidence that deaths occur faster (Deng et al., 2020, Linton et al., 2020). As such, we might be overestimating the healthcare resources needed to treat the most critical patients (namely ventilators). Given the limited evidence for outcome-based durations of resource use, we took a more conservative approach, assuming planners would prefer to overestimate resources needs than under-prepare. With treatment, fatality among infected (IFR) is age-dependent (Verity et al., 2020) (Table 2). When hospital capacity is overwhelmed, we assume a 1% increase in the IFR, chosen to approximately double the IFR in Chile, based on the observed reduction in IFR in China after treatment capacity was augmented to meet demand (Zhang Zuqin et al., 2020). We also chose to base our mortality increase for untreated CoVID19 patients on hospital bed availability (versus critical care beds or ventilators) since the vast majority of cases do not require critical care (~90% of cases in Chile). When more data become available, these assumptions can be updated. Finally, we assume when beds become free at overwhelmed hospitals, new admissions are not associated with a patient's potential outcome.

To estimate the impact of interventions on deaths we calculate infection fatality rates with and without interventions and the number of estimated deaths averted, as the difference in our estimates of cumulative deaths with and without interventions.

*2.3. Illustrative Scenarios and Sensitivity Analyses*

To illustrate the model, we estimated the impact of implementing three intervention strategies in three regions of Chile with the most detected cases through April 6, 2020: Región Metropolitana (RM), an urban region with the largest population including the country's capital Santiago, and Araucanía and Ñuble, two of the least dense urban regions in Chile, but which had experienced rapid growth in late March and were reporting treatment capacity was already strained. We implemented the following three intervention strategies (Table 1) in each region, beginning April 1: Strategy 1) Closure of schools and universities and Telework for 8 months; Strategy 2) Case isolation, home quarantine, social distancing of individuals >70 years, and Telework for 8 months; and Strategy 3) Lockdown for 2 months (6 months shorter duration than the other strategies because the social and economic costs of this suppression strategy are not considered sustainable for the long-term). We chose these strategies because they are in use to some degree in all three regions (Ministry of Health, 2020b, 2020c) (Supplementary Material 2).

We conducted two sensitivity analyses to evaluate the effectiveness of varying the implementation of mitigation strategies. First, we evaluated the influence of shortening the duration (by 2, 4, and 6 months) of mitigation strategies by which successfully reduced our healthcare demand estimates to within the range of treatment capacity. This analysis was chosen since policymakers may be pressured to lift mitigation strategies as early as possible due to their social disruption and economic costs. Then we evaluated the impact of combining the Lockdown strategy with all other strategies, so that when the Lockdown strategy ends, another begins and lasts 6 months. This analysis is intended to address the potential for the outbreak to rebound in the absence of an intervention after a lockdown is lifted (Ferguson et al., 2020, Kissler et al., 2020).

**3. Results**

*3.1. Infections and Deaths without intervention*

We estimate in the absence of any interventions, 5,682,168 to 6,592,016 infections would occur over the course of the epidemic period modeled (5/5 – 12/31/20) in RM, 766,015 to 889,054 infections in Araucanía, and 384,509 to 446,285 infections in Ñuble (Figure 1 and Supplementary Material S2). These projected counts reflect the possibility that 80% to 93% of the populations in these regions may be infected in the absence of any control measures or changes in individual behaviors. Under such a scenario, the number of deaths is projected to be between 106,558 to 125,373 in RM (1.9% IFR), 13,860 to 16,378 deaths in Araucanía (1.8% IFR), and 7,247 to 8,520 deaths in Ñuble (1.9% IFR).

*3.2. Hospital Resource Demands with and without interventions*

Without intervening to control the outbreak, demands for all three of the healthcare resources evaluated by our model are projected to exceed capacity sometime in June in RM (Figure 1 and Supplementary Material S2), and peak sometime between the end of July and mid-August. Araucanía and Ñuble are projected to exceed capacity in July and peak sometime in August or September (approximately one month after RM on both metrics). The degree to which demand is projected to exceed supplies differs by region. In RM, peak demand across all resources is 6 to 18 times the projected maximum supplies available. The situation is similar in Araucanía and Ñuble for hospital beds and ventilators, but is more dire for ICU beds: in both regions the unmitigated peak ICU bed demand is between 13 and 47 times the supply.

Among the two mitigation strategies we evaluated (versus the suppression-type Lockdown strategy), Strategy 2 reduced the burden on the healthcare system the most. In RM, compared to the no intervention scenario, this strategy reduced peak hospital bed occupancy demands by a range of 16,024 to 57,225 (35.4-69.2%), ICU bed occupancy between 2,945 to 8,270 beds (44.8-69.0%), and the number of ventilators needed by 1,572 to 4,237 (47.1-69.3%). This strategy would also push the peak demand for healthcare resources back between 7 and 27 weeks, affording policymakers more time to plan or acquire more capacity. Greater percent reductions but similar delays in the peak demand were observed for Araucanía and Ñuble (Supplementary Material S2).

Our results suggest that this strategy can ease the demands for healthcare to levels below projected capacity constraints when the effectiveness of this strategy is at the higher end of our assumed range (i.e. reductions in $R_0$ approach 47.7%) (Table 1, Figure 1).

For policymakers willing to consider more restrictive measures, our results for the Lockdown strategy, suggest it is an incredibly effective strategy, even for its short duration. The pandemic is quickly subdued and remains so for the duration of the lockdown period in all three regions, with the numbers of cases in treatment remaining relatively flat at levels well below treatment capacity. However, once the lockdown is lifted the number of infected begins to rise again, resulting in demand curves similar in size to the no-intervention scenario, but peaking later: sometime between mid- August and late September (Figure 3, panel A).

FIGURE 1 ABOUT HERE

*3.3. Deaths averted with intervention*

Based on the projected capacity to treat COVID patients in each region, the number of deaths resulting from patients being unable to obtain healthcare was 53,515 to 63,836 in RM, 7,130 to 8,567 in Araucanía, and 3,657 to 4,354 in Ñuble. With Strategy 2, the estimated number of deaths averted in RM ranged from 39,006 to 79,233 (36.6-63.2%), between 4,885 and 12,622 in Araucanía (35.2-77.1%), and 2,018 to 6,742 in Ñuble (27.8-79.1%) (Table 4). Lockdown eliminates between 99.8% and 99.9% of deaths in all three regions during the lockdown period, but deaths rise afterwards with the subsequent rebound of transmission.

*3.4. Sensitivity Analyses*

Figure 2 depicts the effects of shortening the duration of intervention Strategy 2 on hospital bed occupancy demands in RM to two (A), four (B), six (C) months of implementation versus our initial eight month (D) duration. Similar to our baseline results for Strategy 3 (Lockdown), these results show that effectiveness of interventions depends upon the length of time they overlap the epidemic period. Specifically, if too many susceptible individuals remain (i.e., insufficient herd immunity) at the time interventions are lifted, transmission will return. Even when interventions are less effective (R is higher), if the intervention remains in place past peak demand, the resulting outbreak may be smaller than when the same strategy is more effective (R is lower) but lifted prior to peak demand (Figure 2, Panel C).

FIGURE 3 ABOUT HERE

Figure 3 illustrates the results for combining a Lockdown suppression strategy with subsequent mitigation strategies in RM. The benefit of this approach are an additional one to two months delay in peak demand timing beyond delays afforded by each of the mitigation strategies on their own. This approach, however, has no effect on the amount of demand (i.e. peak demand is similar to the mitigation strategy without lockdown).

FIGURE 3 ABOUT HERE

## 4. Discussion

In the absence of immunization, our illustrative results suggest the number of severely ill patients could overwhelm treatment capacity as early as late May to mid-June in all three regions of Chile we evaluated. Our projections also suggest that with immediate aggressive action to implement several combinations of interventions the current amount of hospital beds and critical care beds may be sufficient. In specific circumstances, regional authorities may find it easier to augment their current capacity (e.g. ventilators in Ñuble) along with some mitigation strategies to meet demand versus strictly burdening society with disruptive mitigations. Policymakers should be aware, however, that our results indicate that more effective intervention strategies at temporarily suppressing transmission can also result in larger epidemics upon lifting the strategy than less effective, longer-lasting strategies (in the absence of vaccine and changes in individual behavior). As such, it may be necessary to keep societal-wide interventions in place, or intermittently start and stop them again based on active monitoring of cases counts and treatment capacity, until a vaccine or treatment that can be administered outside of the hospital setting are available.

While our projections are reasonable estimates for how the pandemic may play out given our current understanding of SARS-CoV-2, they should not be considered as forecasts of what will occur. This is due to the uncertainty in our understanding of an outbreak that is still unfolding, the application of experiences of other countries to Chile (such as case severity, resource use by non-COVID patients, intervention effectiveness, compliance over time), simplifying assumptions (such as homogenous mixing), and case surveillance uncertainty. We assumed homogenous mixing to make implementing our model in a spreadsheet more tractable, but as a result, do not reflect potential important variabilities in contact patterns stemming from population social and spatial structures or behavior differences that can affect population disease dynamics. Since obtaining accurate data regarding contact patterns during an ongoing outbreak is challenging and because these patterns may evolve with the outbreak we chose to focus on producing the simplest useful model. To address case surveillance uncertainty, users can scale upwards or downwards their case count inputs occurring over the prior two weeks based on perceived underreporting or overreporting and examine the influence on outputs (as we did in our illustrative scenario). Similarly, all assumptions and sources are explicitly presented in the tool, and all can be readily modified by the user to reflect their interests and as new information comes to light. Therefore, users should consider the value of this tool as its ability to support the evaluation of relative differences in results associated with "what-if" scenarios.

COVID Surge has other limitations worth noting. Our estimates of beds and ventilators needed and the number of deaths averted also depends on associated resources not modeled here. Such resources include trained staff (respiratory therapists, nurses, and physicians) for the successful clinical management of hospitalized and ventilated patients and ancillary supplies associated with a bed or ventilator (e.g. electric circuits, oxygen, etc.). Furthermore, these resources may be impacted by the pandemic itself: staff absenteeism due to illnesses (Wu and McGoogan, 2020) and supply-chain disruptions in personal protective equipment (PPE) for healthcare personnel may further exacerbate the situation. The effects of seasonality on the transmission dynamics of COVID19 remains unclear, but the transmission of similar respiratory illnesses (e.g., influenza, syncytial virus) peaks in the wintertime (Lipsitch and Viboud, 2009, Shaman and Kohn, 2009). If COVID19 exhibits similar seasonality, or patients with these other illnesses place additional demands on the healthcare system, there may be even fewer resources available to treat COVID19 patients at the epidemic's peak. Finally, we do not differentiate between specialized pediatric and non-pediatric resources. While this is justifiable because the current pandemic does not appear to pose a great enough risk to children to overwhelm pediatric healthcare capacity (Riou et al., 2020, Verity et al., 2020, Wu et al., 2020), users of the tool should take note of this meaningful difference when inputting resource amounts.

*4.1. Conclusions*

Our model provides decision makers with the ability to examine the impacts of the current COVID-19 pandemic in their jurisdictions and evaluate the effects of various social-distancing mitigation strategies and augmenting treatment capacity on morbidity and mortality. The results of our illustrative scenario underscore the need for policymakers to take immediate and aggressive actions, and if they do so, substantial morbidity and mortality may be averted. As more data become available (e.g. new treatments or healthcare capacity is augmented) and the pandemic evolves (e.g. COVID case counts), our tool permits rapid updating of results applicable for making decisions.

**Acknowledgments:** The authors thank Shannon Self-Brown for feedback on the manuscript and Marcelo Arenas and Rafael Araos for thoughtful comments and suggestions about applicability of the tool in the Chilean context.

**Supplementary Material 1 (S1):** This is the modeling tool. It contains all data and calculations used in the analyses. There are an English and Spanish version of the tool available for download:
English: https://publichealth.gsu.edu/files/2020/04/Supplementary-Material-S1_Model-v1.0.xlsx
Spanish: https://publichealth.gsu.edu/files/2020/05/SupplMatS1_Model_Espanol_v1.0.xlsx

# References

Chen J, Qi T, Liu L, Ling Y, Qian Z, Li T, et al. Clinical progression of patients with COVID-19 in Shanghai, China. J Infect 2020.

Deng X, Yang J, Wang W, Wang X, Zhou J, Chen Z, et al. Case fatality risk of novel coronavirus diseases 2019 in China. medRxiv 2020.

Dong E, Du H, Gardner L. An interactive web-based dashboard to track COVID-19 in real time. Lancet Infect Dis 2020;online first.

Ferguson N, Laydon D, Nedjati Gilani G, Imai N, Ainslie K, Baguelin M, et al. Report 9: Impact of non-pharmaceutical interventions (NPIs) to reduce COVID19 mortality and healthcare demand. 2020.

Flaxman S, Mishra S, Gandy A. Estimating the number of infections and the impact of non-pharmaceutical interventions on COVID-19 in 11 European countries. Imperial College COVID-19 Response Team 2020;30.

Healthcare P. COVID-19 Hospital Impact Model for Epidemics (CHIME); 2020. Available from: https://penn-chime.phl.io/. [Accessed 3/31/20.

Henderson M. Covid Act Now. 2020. p. https://www.covidactnow.org/.

ICNARC. ICNARC report on COVID-19 in critical care. 27 March 2020. London: Intensive Care National Audit & Research Centre; 2020.

Instituto Nacional de Estadísticas. Estimaciones y proyecciones de la población de Chile 1992-2050; 2017. Available from: https://www.censo2017.cl/. [Accessed April 2 2020].

Kissler SM, Tedijanto C, Lipsitch M, Grad Y. Social distancing strategies for curbing the COVID-19 epidemic. medRxiv 2020.

Latorre R, Sandoval G. El mapa actualizado de las camas de hospitales en Chile. La Tercera. Santiago, Chile2020.

Lauer SA, Grantz KH, Bi Q, Jones FK, Zheng Q, Meredith HR, et al. The incubation period of coronavirus disease 2019 (COVID-19) from publicly reported confirmed cases: estimation and application. Annals of internal medicine 2020.

Li G, De Clercq E. Therapeutic options for the 2019 novel coronavirus (2019-nCoV). Nature Reviews Drug Discovery 2020;19:149-50.

Li Q, Guan X, Wu P, Wang X, Zhou L, Tong Y, et al. Early transmission dynamics in Wuhan, China, of novel coronavirus–infected pneumonia. N Engl J Med 2020.

Linton NM, Kobayashi T, Yang Y, Hayashi K, Akhmetzhanov AR, Jung S-m, et al. Incubation period and other epidemiological characteristics of 2019 novel coronavirus infections with right truncation: a statistical analysis of publicly available case data. Journal of clinical medicine 2020;9(2):538.

Lipsitch M, Viboud C. Influenza seasonality: lifting the fog. Proc Natl Acad Sci 2009;106(10):3645-6.

Meltzer MI, Patel A, Ajao A, Nystrom SV, Koonin LM. Estimates of the demand for mechanical ventilation in the United States during an influenza pandemic. Clin Infect Dis 2015;60(suppl_1):S52-S7.

Ministry of Health. Cifras Oficiales COVID-19; 2020a. Available from: https://www.gob.cl/coronavirus/cifrasoficiales/. [Accessed April 2 2020].

Ministry of Health. Dispone medidas sanitarias que indica por brote de COVID-19. Norms 1143498, 1143591, 1746958. In: Ministry of Health Chile, editor. Santiago: Biblioteca del Congreso Nacional de Chile; 2020b.

Ministry of Health. Dispone régimen especial de cumplimiento de jornada laboral y flexibilidad horaria por brote de coronavirus (COVID-19). Norm 1143629; 2020c. Available from: https://www.leychile.cl/N?i=1143629&f=2020-03-20&p=.

Nature. First vaccine clinical trials begin in the United States; 2020. Available from: https://www.nature.com/articles/d41586-020-00154-w. [Accessed March 17 2020].

OECD. Health at a Glance 2019: OECD Indicators. Paris: OECD Publishing, 2019.

Riou J, Althaus CL. Pattern of early human-to-human transmission of Wuhan 2019 novel coronavirus (2019-nCoV), December 2019 to January 2020. Eurosurveillance 2020;25(4).

Riou J, Hauser A, Counotte MJ, Althaus CL. Adjusted age-specific case fatality ratio during the COVID-19 epidemic in Hubei, China, January and February 2020. medRxiv 2020.

Shaman J, Kohn M. Absolute humidity modulates influenza survival, transmission, and seasonality. Proc Natl Acad Sci 2009;106(9):3243-8.

Verity R, Okell LC, Dorigatti I, Winskill P, Whittaker C, Imai N, et al. Estimates of the severity of coronavirus disease 2019: a model-based analysis. Lancet Infect Dis 2020.

Walker PG, Whittaker C, Watson O, Baguelin M, Ainslie K, Bhatia S, et al. The Global Impact of COVID-19 and Strategies for Mitigation and Suppression. On behalf of the imperial college covid-19 response team, Imperial College of London 2020.

Wang C, Liu L, Hao X, Guo H, Wang Q, Huang J, et al. Evolving Epidemiology and Impact of Non-pharmaceutical Interventions on the Outbreak of Coronavirus Disease 2019 in Wuhan, China. medRxiv 2020.

Wang D, Hu B, Hu C, Zhu F, Liu X, Zhang J, et al. Clinical characteristics of 138 hospitalized patients with 2019 novel coronavirus–infected pneumonia in Wuhan, China. Jama 2020.

Willem L, Hoang TV, Funk S, Coletti P, Beutels P, Hens N. SOCRATES: An online tool leveraging a social contact data sharing initiative to assess mitigation strategies for COVID-19. medRxiv 2020:2020.03.03.20030627.

World Health Organization. Coronavirus Disease (COVID-19) Outbreak; 2020. Available from: https://www.who.int/emergencies/diseases/novel-coronavirus-2019. [Accessed 17 March 2020].

Wu JT, Leung K, Bushman M, Kishore N, Niehus R, de Salazar PM, et al. Estimating clinical severity of COVID-19 from the transmission dynamics in Wuhan, China. Nat Med 2020.

Wu Z, McGoogan JM. Characteristics of and important lessons from the coronavirus disease 2019 (COVID-19) outbreak in China: summary of a report of 72 314 cases from the Chinese Center for Disease Control and Prevention. Jama 2020.

Wunsch H, Wagner J, Herlim M, Chong D, Kramer A, Halpern SD. ICU occupancy and mechanical ventilator use in the United States. Critical care medicine 2013;41(12).

Yang X, Yu Y, Xu J, Shu H, Xia Ja, Liu H, et al. Clinical course and outcomes of critically ill patients with SARS-CoV-2 pneumonia in Wuhan, China: a single-centered, retrospective, observational study. The Lancet Respiratory Medicine 2020.

You C, Deng Y, Hu W, Sun J, Lin Q, Zhou F, et al. Estimation of the Time-Varying Reproduction Number of COVID-19 Outbreak in China. medRxiv 2020:2020.02.08.20021253.

Zhang G, Hu C, Luo L, Fang F, Chen Y, Li J, et al. Clinical features and outcomes of 221 patients with COVID-19 in Wuhan, China. medRxiv 2020.

Zhang Z, Yao W, Wang Y, Long C, Fu X. Wuhan and Hubei COVID-19 mortality analysis reveals the critical role of timely supply of medical resources. medRxiv 2020.

Zhu N, Zhang D, Wang W, Li X, Yang B, Song J, et al. A novel coronavirus from patients with pneumonia in China, 2019. N Engl J Med 2020.

**Table 1.** Intervention Strategies and Effects on Onward Transmission

| Strategy Name (Strategy Type) | Description | Reduction in $R_0$[a] Low[b] | High[b] |
|---|---|---|---|
| **Case isolation** (mitigation) | Symptomatic cases stay at home for 7 days, reducing non-household contacts during this period. Household contacts remain unchanged. | 15.8% | 18.6% |
| **Closing Schools and Universities + Telework** (mitigation) | <u>Closing Schools/Universities</u>: Physical closure of all schools and universities (or move to virtual learning environment). Assumes some increase in contacts in the household and the community during the closure, partially offsetting reductions in transmission at schools and universities.<br><br><u>Telework</u>: All government switches to telework to the maximum extent possible and private businesses are encouraged to telework, resulting in 50% of the working population teleworking. | 15.8% | 16.8% |
| **Case isolation + Household quarantine** (mitigation) | <u>Case isolation</u>: same as above<br><br><u>Household quarantine</u>: Following identification of a symptomatic case in the household, all household members voluntarily remain at home for 14 days. Increased transmission between household members during the quarantine period will partially offset transmission reductions in the community. | 25.4% | 30.0% |
| **Case isolation + Household quarantine + Social distancing of >70s + Telework** (mitigation) | <u>Case isolation</u>: same as above<br><br><u>Household quarantine</u>: same as above<br><br><u>Social Distancing of >70s</u>: Reduce contacts among older individuals (>70 years of age) because of their increased risk for severe outcomes and healthcare resource requirements. These individuals reduce contacts outside the home by 50%.<br><br><u>Telework</u>: same as above | 41.9% | 47.7% |
| **Lockdown** (suppression) | Population-wide social distancing by forced quarantine of all households and workplaces, and border closed to travel. Only essential outings from the home are permitted (e.g. food/supplies purchases) and for employees working at businesses deemed essential for continued operation. | 57.7% | 68.2% |

**Notes**

[a] $R_0$ = *basic reproduction number*. It represents the average number of people who will be infected by any given infected person at the early stages of disease spread when there are no control measures.

[b] High and Low values of the reduction in transmission associated with each strategy were used to account for uncertainty in societal compliance and strategy effectiveness. These reductions were based on equivalent reductions in Critical Care Bed Occupancy published in Ferguson et al. (2020) (Supplementary Material S2). We added 10 percentage points to reduction values for strategies including telework, based on Willem et al. (2020).

**Table 2.** Risk of Healthcare Use and Outcomes Among Infected

| Age group | % Infected, Hospitalized[a] | % of Hospitalized, Admitted to ICU[a] | % ICU patients needing ventilation[b] | Infection Fatality Ratio (IFR)[a] | Fatality Increase if Demand>Capacity[c] |
|---|---|---|---|---|---|
| 0-9 | 0.01% | 5.0% | 63.2% | 0.002% | 1.000% |
| 10-19 | 0.04% | 5.0% | 63.2% | 0.006% | 1.000% |
| 20-29 | 1.10% | 5.0% | 63.2% | 0.030% | 1.000% |
| 30-39 | 3.40% | 5.0% | 63.2% | 0.080% | 1.000% |
| 40-49 | 4.30% | 6.3% | 63.2% | 0.150% | 1.000% |
| 50-59 | 8.20% | 12.2% | 63.2% | 0.600% | 1.000% |
| 60-69 | 11.80% | 27.4% | 63.2% | 2.200% | 1.000% |
| 70-79 | 16.60% | 43.2% | 63.2% | 5.100% | 1.000% |
| 80+ | 18.40% | 70.9% | 63.2% | 9.300% | 1.000% |

**Notes**
[a] Verity et al. (2020)
[b] Based on ICNARC (2020). Alternative estimates include 60% (Meltzer et al., 2015) and 71.1% (Yang et al., 2020).
[c] Percentage points increase in fatality when hospitals are overwhelmed. We assumed a 1% increase in the IFR to approximately double the population-weighted age-based IFR in Chile, based on data from COVID19 in China (Zhang Zuqin et al., 2020)

**Table 3.** Model Inputs by Region for all Illustrative Scenarios

|  | Region | | | Source |
|---|---|---|---|---|
|  | Metropolitana | Araucanía | Ñuble |  |
| **Population**[a] | 7,112,808 | 957,224 | 480,609 | INE(2017) |
| **COVID-19 reported cases**[b] | | | | |
| Cumulative | 20,590 | 1,907 | 1,107 | Ministry of Health (2020a) |
| 2 weeks through 05/04/20 | 12,487 | 364 | 133 | Ministry of Health (2020a) |
| **$R_0$** | 2.0 - 2.8 | 2.0 - 2.8 | 2.0 - 2.8 | Riou et al, Li et al (2020) |
| **Intervention Strategy** | | | | |
| School closures, telework | 4/1-12/1/20 | 4/1-12/1/20 | 4/1-12/1/20 | Assumed |
| Case isolation, home quarantine, social distancing>70, telework | 4/1-12/1/20 | 4/1-12/1/20 | 4/1-12/1/20 | Assumed |
| Lockdown | 4/1-6/1/20 | 4/1-6/1/20 | 4/1-6/1/20 | Assumed |
| **Disease severity** | | | | |
| Infected who are hospitalized[c] (%) | 4.5% | 4.8% | 5.1% | Verity et al. (2020) |
| Hospitalized, admitted to ICU[c] (%) | 11.4% | 12.2% | 12.7% | Verity et al. (2020) |
| Infection Fatality rate[c] (%) | 0.8% | 0.9% | 0.9% | Verity et al. (2020) |
| ICU patients needing ventilator (%) | 63.2% | 63.2% | 63.2% | ICNARC (2020) |
| **Healthcare resources**[d] | | | | |
| Hospital (non-ICU) beds | 18,522 | 2,671 | 1,010 | Latorre et al.(2020) |
| In-use by Non-COVID Patients (%) | 71% | 71% | 71% | OECD (2019) |
| In-use by COVID Patients (%)[e] | 3% | 3% | 3% | Ministry of Health (2020a) |
| Critical Care Beds | 2,326 | 215 | 60 | Latorre et al.(2020) |
| In-use by Non-COVID Patients (%) | 71% | 71% | 71% | OECD (2019) |
| In-use by COVID Patients (%)[e] | 14% | 14% | 14% | Ministry of Health (2020a) |
| Ventilators | 867 | 80 | 22 | Latorre et al.(2020) |
| In-use by Non-COVID Patients (%)[f] | 40% | 40% | 40% | Wunsch et al. (2013) |
| In-use by COVID Patients (%)[g] | 19% | 19% | 19% | Assumed |

**Notes**
[a] Population distributed by age groups are shown in the Supplementary Material S2, based on INE's Housing and Population Census 2017 (Instituto Nacional de Estadísticas, 2017)
[b] Scaled counts to account for assumed 40% under-reporting in reported cases (based on 60% reported by Wang et al. (2020) minus 20% to account for improvements in case-detection in Chile since the outbreak's start).
[c] Estimates differ by region due to age structure of the populations (Supplementary Material S2).
[d] All beds available in the healthcare system, from public and private hospitals, are now part of the "Sistema Integrado COVID-19" under the centralized administration of the Ministry of Health. An intensive care bed (ICU) consists of a cot with a monitor, healthcare professionals and medication to treat patient. Some have a mechanical ventilator. There are an estimated 1,847 mechanical ventilators; 850 currently available and 997 were acquired in January 2020.(Latorre and Sandoval, 2020) We assumed the distribution of mechanical ventilators was proportional to the number of critical beds in each region: Metropolitana,47.0%; Araucanía, 4.3%; Ñuble, 1.2%. (Supplementary Material S2)
[e] Based on the reported number hospitalized in "basic beds" (1,216) and in "critical care beds" (699) in all of Chile by the Ministry of Health as of May 4, 2020 out of the total existing beds nationally in March 2020 plus anticipated beds being added to expand pandemic treatment capacity: 41,706 and 4,954, respectively. Latorre et al.(2020)
[f] Availability of mechanical ventilators was based on a three-year study of 97 ICUs in the US.(Wunsch et al., 2013)
[g] Calculated by applying the % ICU patients needing ventilation (Table 2) to the number of COVID patients in critical care beds (see note e) and dividing the result by the total ventilators in Chile (see note f).

**Table 4.** Deaths averted by each intervention strategy and region (compared to deaths without intervention) between May 5, 2020 and December 31, 2020

| Intervention Strategy[a] | Metropolitana | Araucanía | Ñuble |
|---|---|---|---|
| **Strategy 1: School closures, telework** | 7,612 – 20,725 (7.1 – 16.5%) | 1,019 – 2,707 (7.4 – 16.5%) | 518 – 1,288 (7.1 – 15.1%) |
| **Strategy 2: Case isolation, home quarantine, social distancing>70, telework** | 39,006 – 79,233 (36.6 – 63.2%) | 4,885 - 12,622 (35.2 - 77.1%) | 2,018 - 6,742 (27.8 - 79.1%) |
| **Strategy 3: Lockdown** [b] | 106,381 – 125,140 (99.8 - 99.8%) | 13,855 – 16,372 (99.9 – 99.9%) | 7,706 – 8,274 (99.9 – 99.9%) |

**Notes**
[a] Implemented per scenarios in Table 3 and assumptions in Tables 1-2.
[b] Values shown are based on deaths during the lockdown period only due to its short duration and the subsequent rise in deaths when lockdown ends (Figure 1).

**Figure 1.** Projected occupancy demands and capacity for hospital (non-ICU) beds in Región Metropolitana with and without intervention.

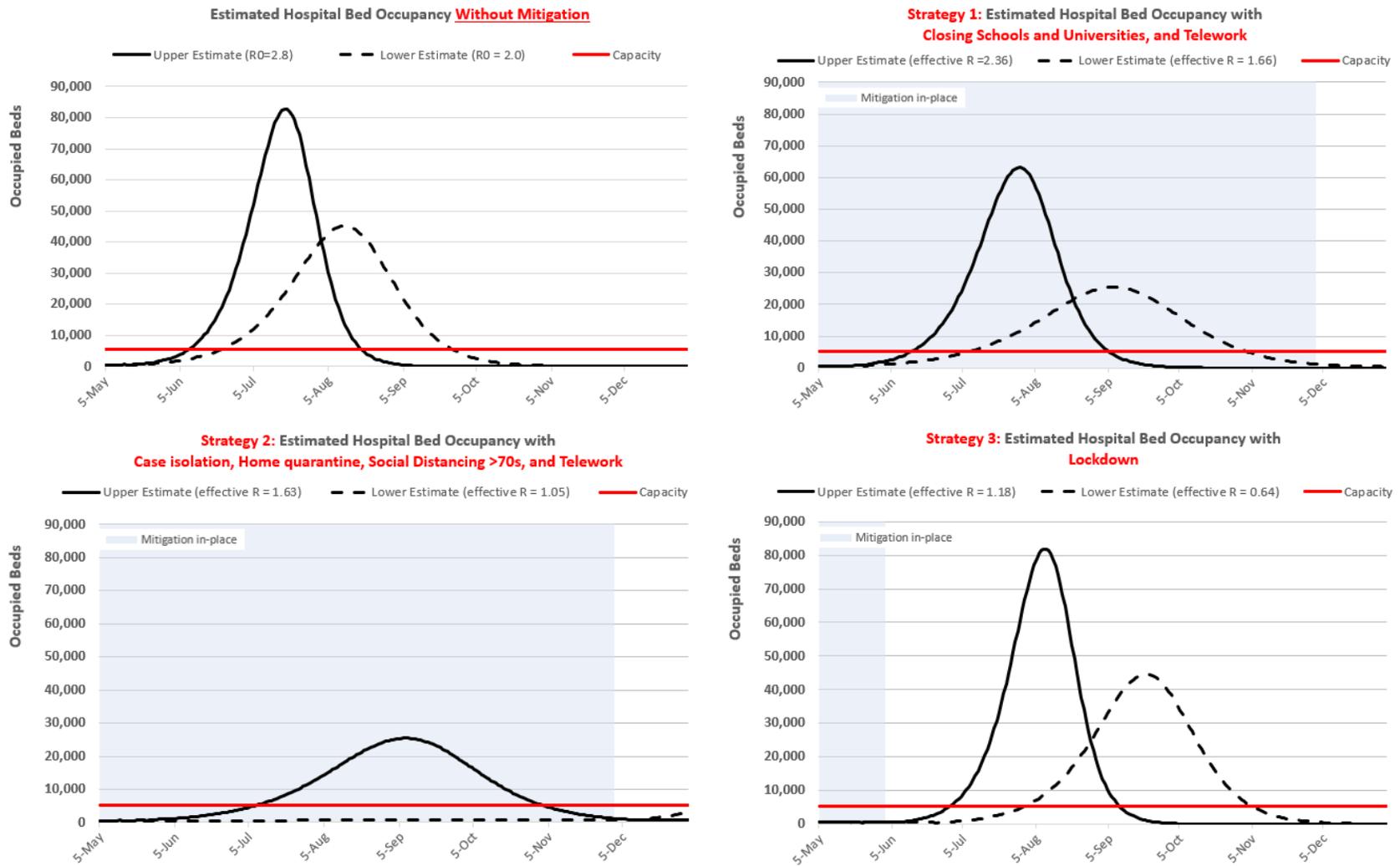

**Notes.** Solid curves: projections using the high estimate for the reproduction number. Dashed curves: projections using the low estimate for the reproduction number. Table 1 contains all reproduction numbers. Horizontal Red line: Hospital bed capacity. Blue shaded region: interventions in place.

**Figure 2.** Sensitivity analysis: Effects of the duration of intervention Strategy 2 (case isolation, home quarantine, social distancing of population >70 years of age, and telework) on hospital bed occupancy demands during the COVID-19 epidemic in Región Metropolitana when maintained for two (A), four (B), six (C), and eight (D) months (and initiated on April 1, 2020).

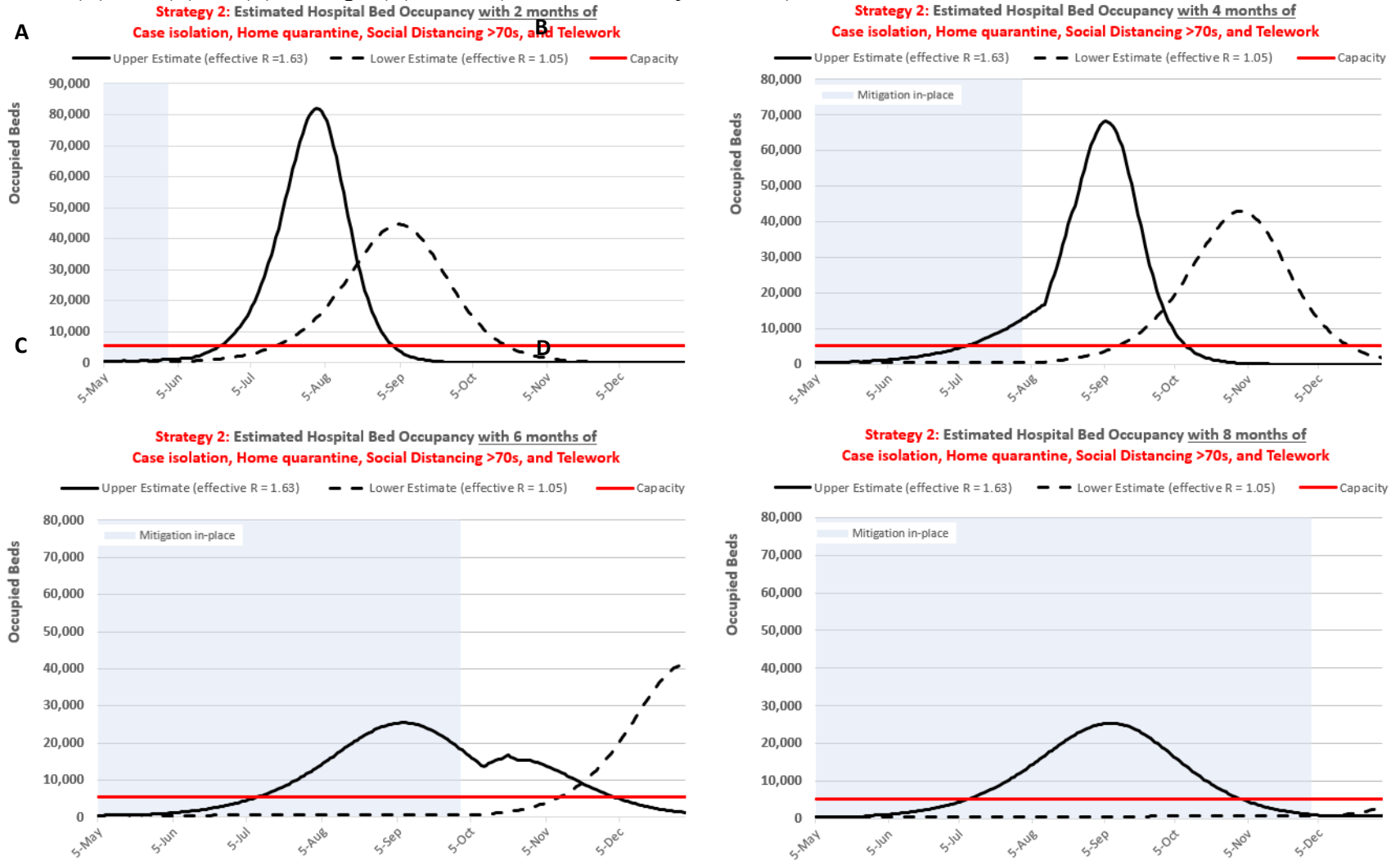

Notes: Solid and dashed curves reflect uncertainty in the effectiveness of intervention strategies (Table 1)

**Figure 3.** Sensitivity Analysis: Effects of a 2 month Lockdown Suppression Strategy alone (A) and followed by various mitigation strategies for 6 months on Hospital Bed Occupancy Demands: Closing Schools and Universities + Telework (B), Case Isolation + Household Quarantine (C), and Case isolation, Household Quarantine, Social Distancing of >70 years of age, and Telework (D)

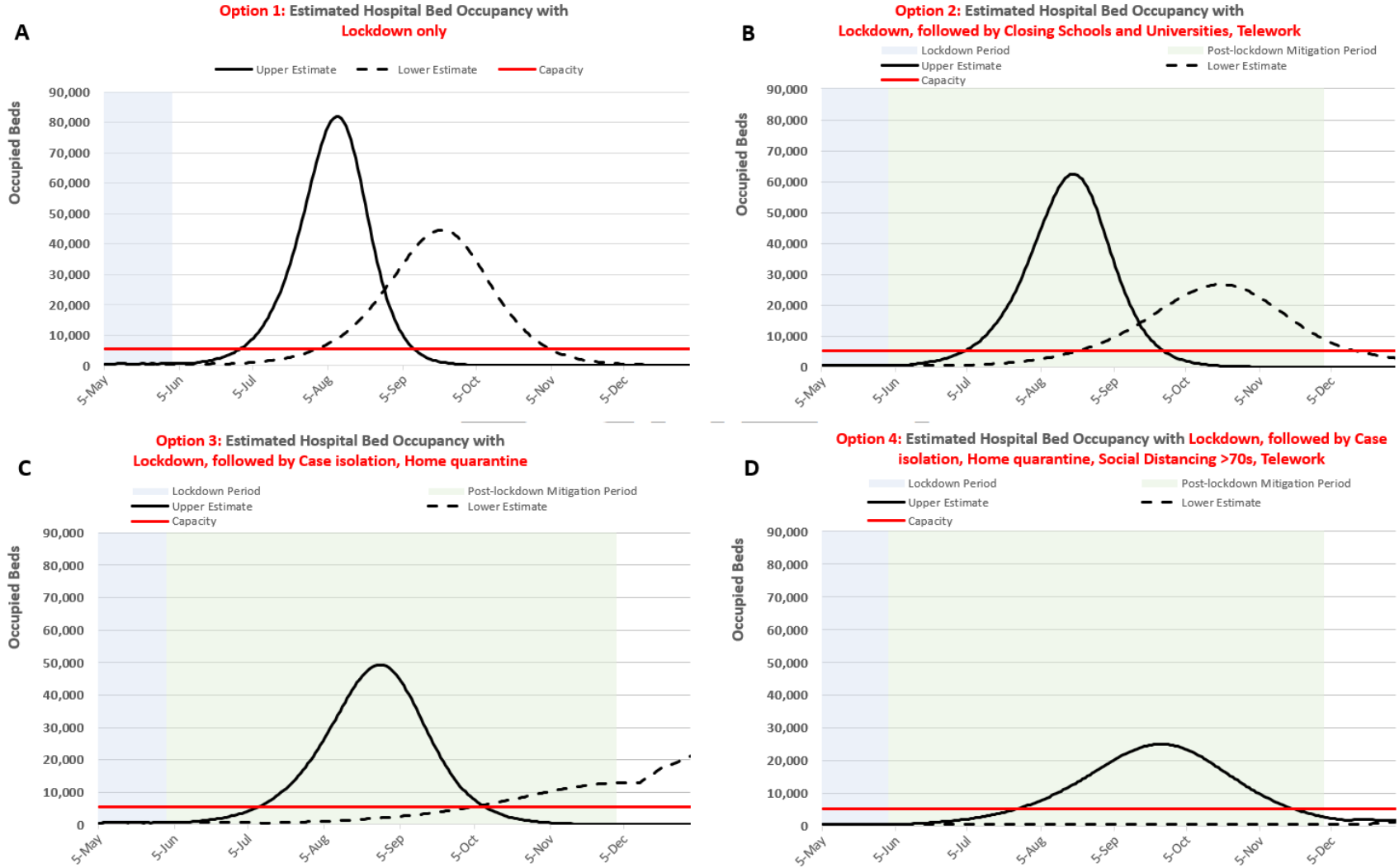

Notes: Solid and dashed curves reflect uncertainty in the effectiveness of intervention strategies during both the Lockdown period and Post-lockdown intervention period per Table 1.

**Supplementary Material 2**

This appendix provides further details on the methods used as well as additional results.

**SEIR Model**

The model consists of individuals who are either *Susceptible* (S), *Infected* but not yet Infectious (E), Infectious (I), Total *Recovered* and Died (D). It projects and tracks the number of individuals moving between these categories every day of the outbreak. Projections begin on the day following the date input by users for the last day of the most recent 2-weeks of cases available. On this date, there are only Susceptible and Infectious individuals. The epidemic then proceeds via a growth and decline process: As the number of susceptible individuals is depleted (once individuals are infected) the spread of the infection slows. Individuals in the Infectious (I) category includes those who are not yet symptomatic (pre-symptomatic) but will become symptomatic, those who are symptomatic, and those who are infectious yet not exhibiting symptoms (asymptomatic). The dynamics are given by the following equations such that on any given day t, the number individuals Susceptible (S), Infectious (I), Recovered (L), and Died (D) are:

$$S_t = S_{t-1} - \left(\frac{R}{\gamma} * S_{t-1} * \frac{I_{t-1}}{N}\right)$$

$$I_t = \sum_{i=t-(1/\gamma)-\kappa-1}^{t-(1/\gamma)-1} \frac{R}{\gamma} * S_{t-1} * \frac{I_{t_i}}{N}$$

$$L_t = \left(\frac{R}{\gamma} * S_{t-(\kappa+1/\gamma)-1} * \frac{I_{t-(1/\gamma)-\kappa-1}}{N}\right)(1-\alpha)$$

$$D_t = \left(\frac{R}{\gamma} * S_{t-(\kappa+1/\gamma)-1} * \frac{I_{t-(1/\gamma)-\kappa-1}}{N}\right)\alpha$$

where:
N is the population size,

R is the number of new infections each infected persons causes with R = $R_0$ when no mitigation strategy is in place and R=$R_e$ when a mitigation strategy is being used, and κ is the number of days needed to become infectious after being infected.

1/γ is the number of days needed to recover (or die) once infectious.

α is the proportion of infected that die (i.e. infection fatality rate (IFR)) with $\alpha = \alpha_1$ when hospitals have capacity to treat and $\alpha = \alpha_2$ when capacity is overwhelmed

Note: the Infected but not yet Infectious state (E) is not calculated each day in our tool (i.e. not given it's own data column), but still contributes to the model by delaying when infected persons begin to contribute to the force of infection.

## Reductions in $R_0$ Associated with Interventions

We chose the reductions in $R_0$ for each intervention strategy by determining the reduction applied to $R_0 = 2.4$ (no intervention) in our model which produced comparable percent declines and delays in peak critical care (ICU) bed occupancy from the "do nothing scenario" observed in Ferguson et al.'s [1] Figure 2. Table S1 shows how these were determined.

Table S1. Summary of values used for determining the reductions in R associated with interventions

| Intervention Strategy | Observed in Figure 2 of Ferguson et. al. | | | | Observed in our model | | | |
|---|---|---|---|---|---|---|---|---|
| | (A) Peak Occupancy | (B) Weeks peak is delayed (compared to "Do Nothing") | (C) Beds Occupied | (D) Reduction in Beds Occupied (compared to "Do Nothing") | (E) Median $R_0$ | (F) Peak Occupancy | (G) Weeks peak is delayed (compared to "Do Nothing") | (H) Reduction in Beds Occupied |
| Do Nothing | 5/19/2020 | -- | 2700 | -- | 2.4 | 19,538* | -- | -- |
| Closing Schools and Universities | 5/26/2020 | 1 | 2400 | 11.1% | 2.25 | 1087 | 1 | 10.9% |
| Case isolation | 6/1/2020 | 2 | 1850 | 31.5% | 1.98 | 831 | 3 | 31.9% |
| Case isolation + household quarentine | 6/7/2020 | 3 | 1300 | 51.9% | 1.74 | 590 | 5 | 51.6% |
| Case isolation, home quarantine, social distancing of >70s | 6/10/2020 | 3 | 920 | 65.9% | 1.57 | 417 | 8 | 65.8% |
| Lockdown | 4/15/2020 | -5 | 5 | 99.8% | 0.9 | 3 | -11 | 99.8% |

**Notes**
* Generated using 2,793 cases through March 31 for all of Chile (per population 17,574,003), and interventions beginning the next day (4/1/20) and continuing through the calendar year
A-C: Estimated from Figure 2 Ferguson, Laydon [1].
D: $(C_{Do\ nothing} - C_{intervention}) / C_{Do\ nothing} * 100$
E: Manipulated manually in our model until columns E and G approximated columns B
F-G: Observed in our model
H: $(F_{Do\ nothing} - F_{intervention}) / F_{Do\ nothing} * 100$

**Interventions Application Timeline in Chile** [2,3]

Closure of all daycares, schools, and universities was mandated across all of Chile on March 16; followed by case isolation, and mandatory home quarantine for CoVID19 patients on March 19, and the implementation of flexible work schedules and telework for government workers began March 20. Social distancing measures across Chile include bans on nursing homes visits (03/16), closures of non-essential business (03/20, e.g., restaurants, pubs, night clubs), night curfews (03/22), and bans on meetings and events ≥50 people (03/24). Additionally, since March 28, two major cities in Araucanía, and seven municipalities in Santiago are under a mandatory lockdown.

**Table S2**. Demographics of the Chilean population in the three study regions

| Age group | Chile | Metropolitana XIII | Ñuble XVI | Araucanía IX |
|---:|---:|---:|---:|---:|
| 0-9 | 2,376,335 | 937,432 | 61,464 | 133,392 |
| 10-19 | 2,392,112 | 933,218 | 66,179 | 137,493 |
| 20-29 | 2,861,972 | 1,238,583 | 66,985 | 144,782 |
| 30-39 | 2,501,414 | 1,066,451 | 60,993 | 124,652 |
| 40-49 | 2,359,266 | 951,497 | 67,450 | 128,716 |
| 50-59 | 2,232,733 | 889,726 | 66,574 | 120,577 |
| 60-69 | 1,499,917 | 579,388 | 46,661 | 84,658 |
| 70-79 | 879,498 | 333,994 | 29,403 | 53,289 |
| 80+ | 470,756 | 182,519 | 14,900 | 29,665 |
| **Urban** | 15,424,263 | 6,849,310 | 333,680 | 678,544 |
| **Rural** | 2,149,740 | 263,498 | 146,929 | 278,680 |
| **Total** | 17,574,003 | 7,112,808 | 480,609 | 957,224 |

**Notes.** Chile has a total of 16 regions. Here we include the three regions that have been more heavily affected by CoVID-19 as of April 5th 2020, since the first case was reported in march 2, as an illustration of the potential uses of the tool. Estimates for all regions have been reported to the Ministry of Health.
**Source:** CENSO 2017.[4]

**Table S3.** Reported cases of COVID-19 by region

| Region | N | Total | Two weeks |
|---|---|---:|---:|
| Metropolitana | XIII | 2350 | 1810 |
| Araucanía | IX | 612 | 553 |
| Ñuble | XVI | 522 | 417 |
| Chile | | 5116 | 4194 |

**Notes.** Total reported CoVID-19 cases as of April 5 2020
Source: Ministry of Health[5]

**Table S4.** Healthcare capacity: basic and intensive care beds by region, public and private hospitals

|  | Basic beds | | | Beds/ 100k | Intensive care | | | Beds/ 100k | Mech. Ventilator |
| --- | --- | --- | --- | --- | --- | --- | --- | --- | --- |
|  | Current | Increase† | Total |  | Current | Increase† | Total |  |  |
| Metropolitana | 16,596 | 1,926 | 18,522 | 260.4 | 1,937 | 389 | 2,326 | 32.7 | 867 |
| Ñuble | 942 | 68 | 1,010 | 210.2 | 38 | 22 | 60 | 12.5 | 22 |
| Araucanía | 2,202 | 469 | 2,671 | 279.0 | 136 | 79 | 215 | 22.5 | 80 |
| Chile | 37,777 | 3,929 | 41,706 | 227.2 | 3,295 | 1,659 | 4,954 | 29.1 | 1,847 |

**Notes.**
†Increase refers to new beds in the health care system as a consequence of CoVID-19 response. All beds available in the healthcare system, from public and private hospitals, are now part of the "Sistema Integrado COVID-19" under the centralized administration of the Ministry of Health. An intensive care bed (ICU) consists of a cot with a monitor, healthcare professionals and medication to treat patient. Some have a mechanical ventilator.

There are an estimated 1,847 mechanical ventilators; 850 currently available and 997 were acquired in January 2020.[6] We assumed the distribution of mechanical ventilators was proportional to the number of critical beds in each region. We assumed 60% of mechanical ventilators would be available based on a three-year study of 97 ICUs in the US, including 226,942 admissions to ICUs.[7]

**Source:** Latorre et al. 2020[6]

## B. ADDITIONAL RESULTS
### B.1.1 Región Metropolitana – Hospital beds

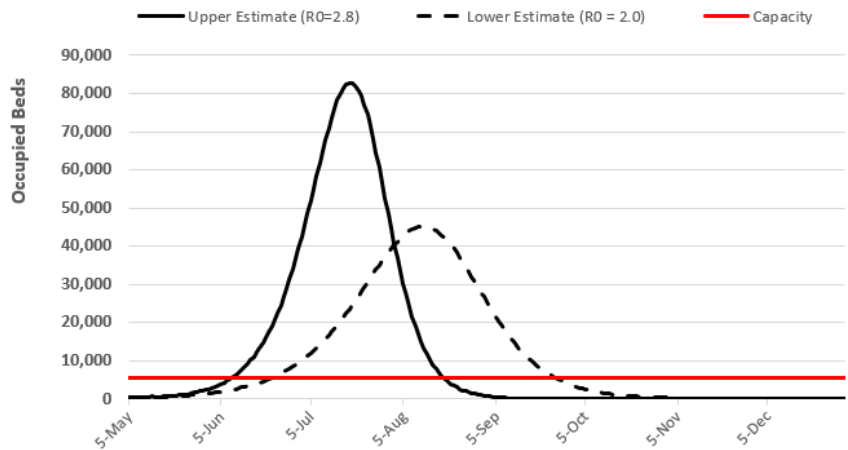

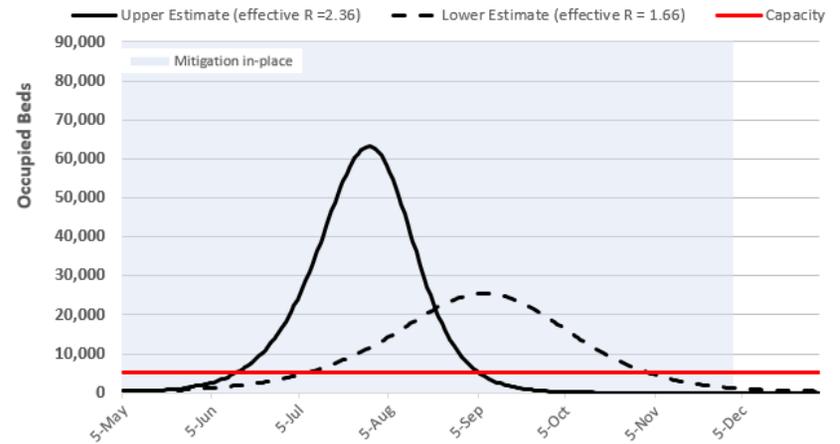

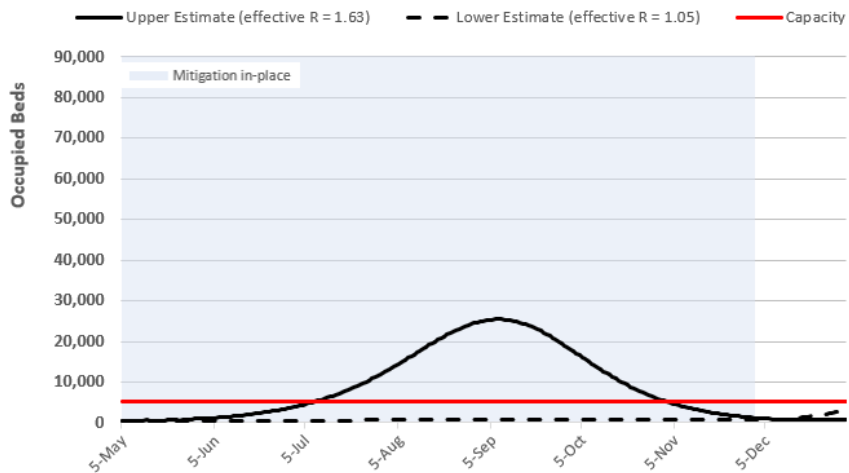

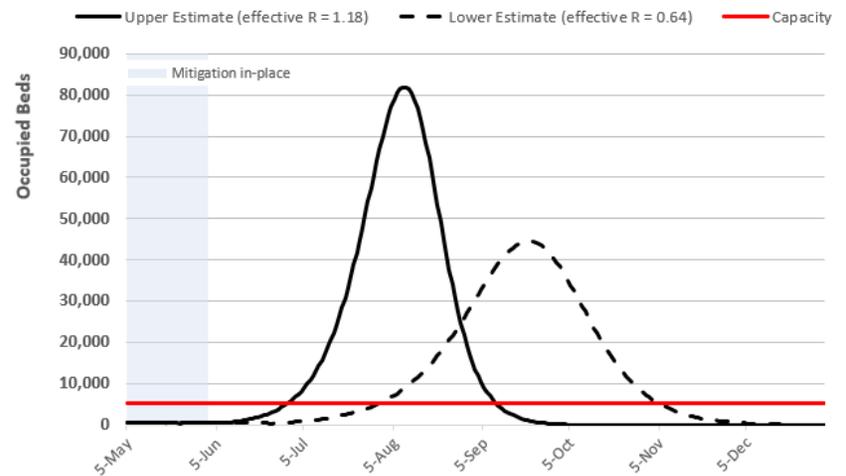

## B.1.2 Región Metropolitana – ICU Beds

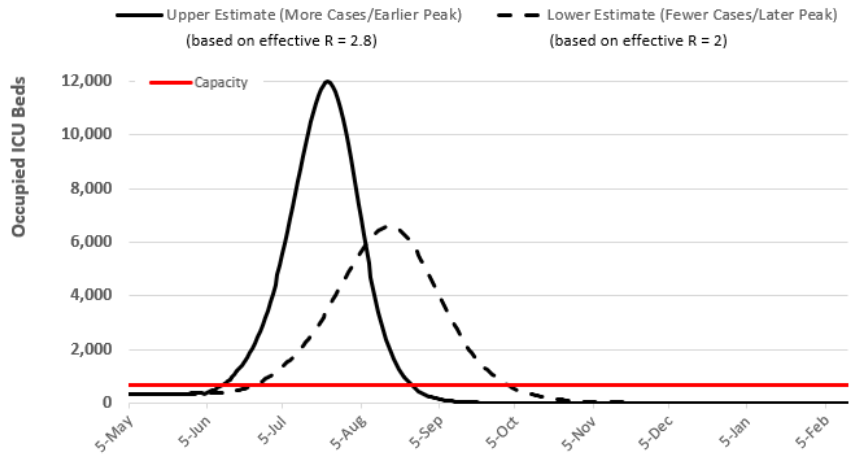
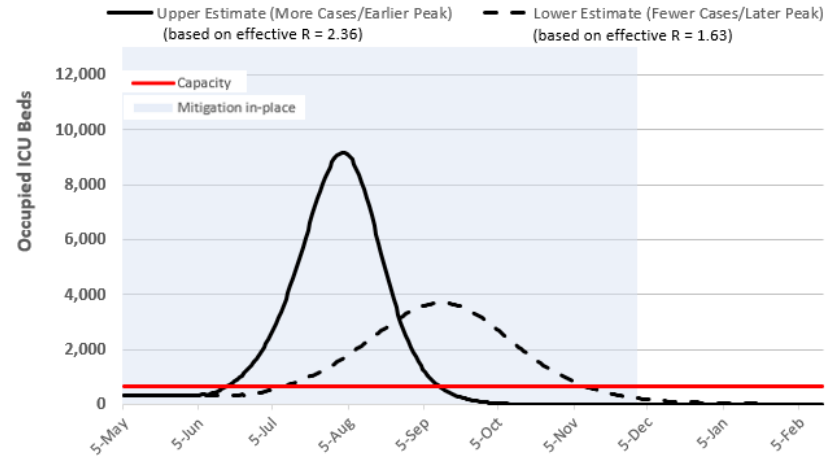
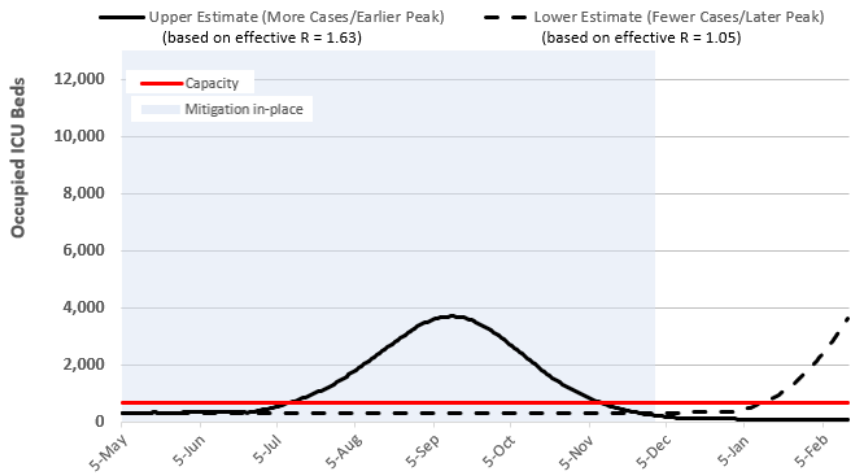
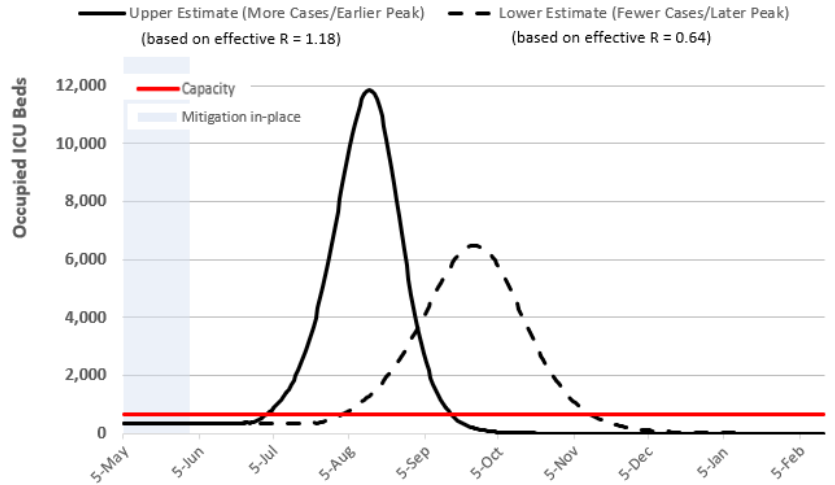

### B.1.3 Región Metropolitana – Ventilators

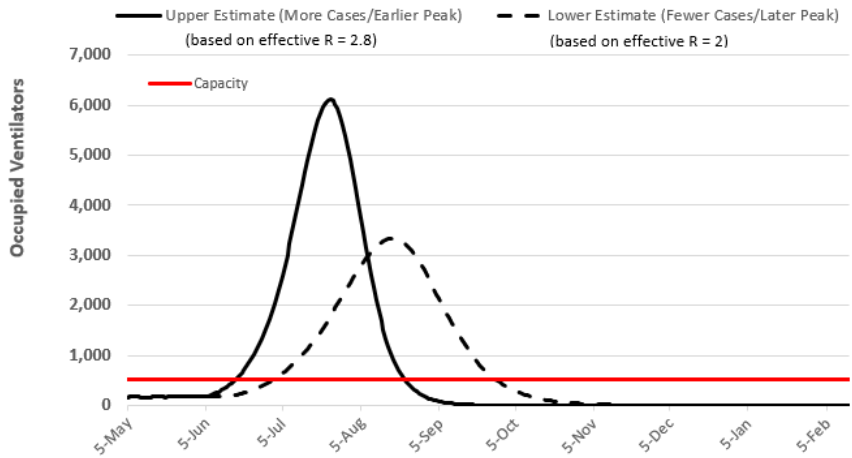
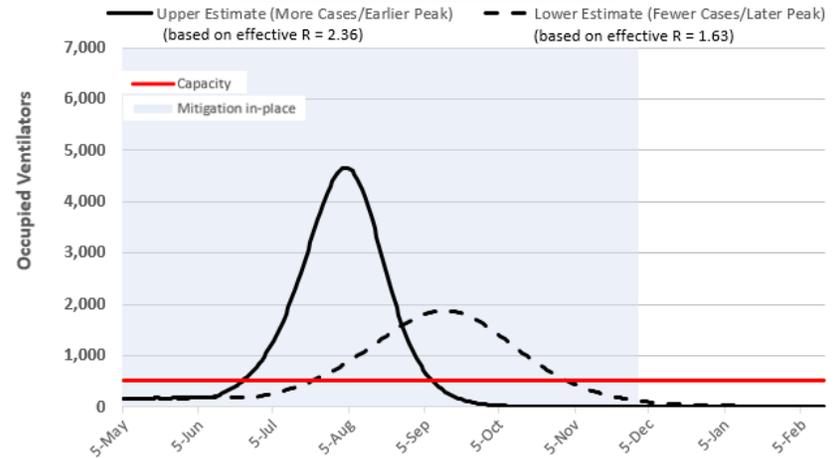
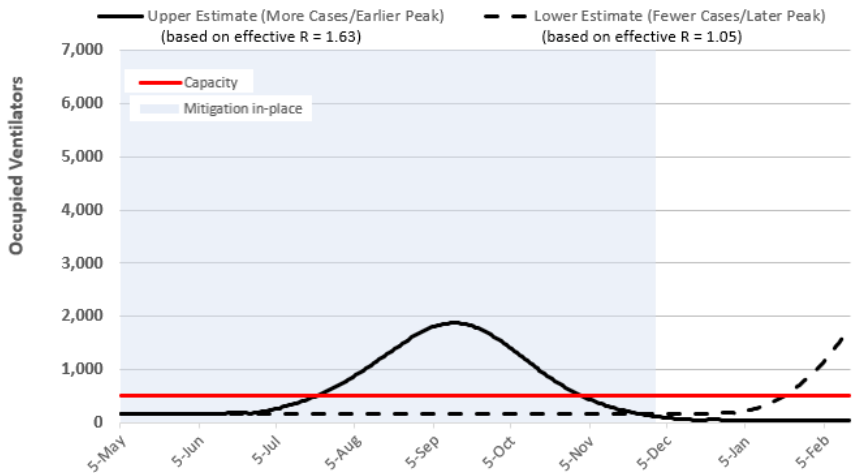
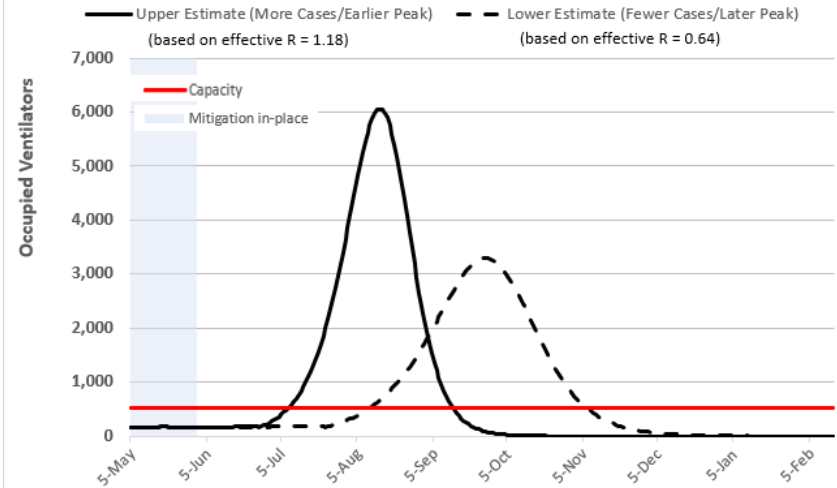

## B.2.1 Región Araucanía - Hospital beds

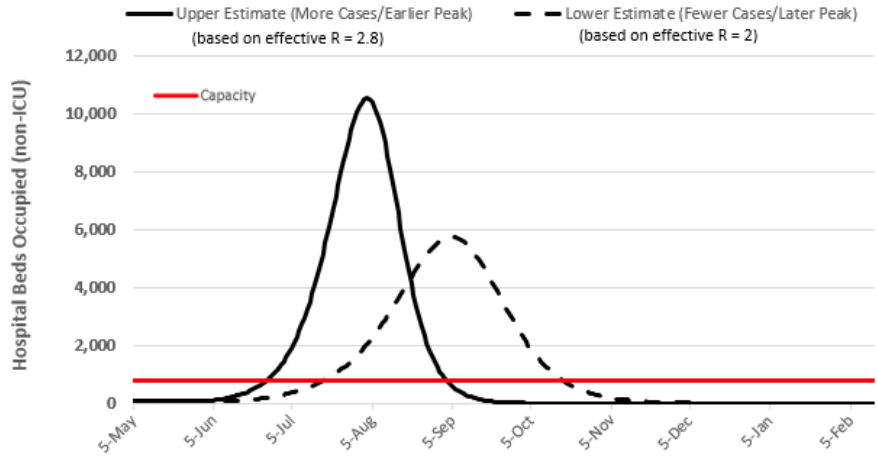
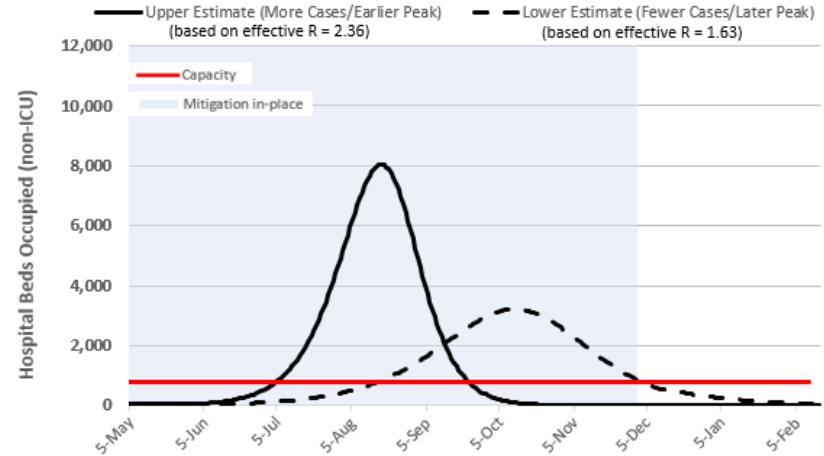
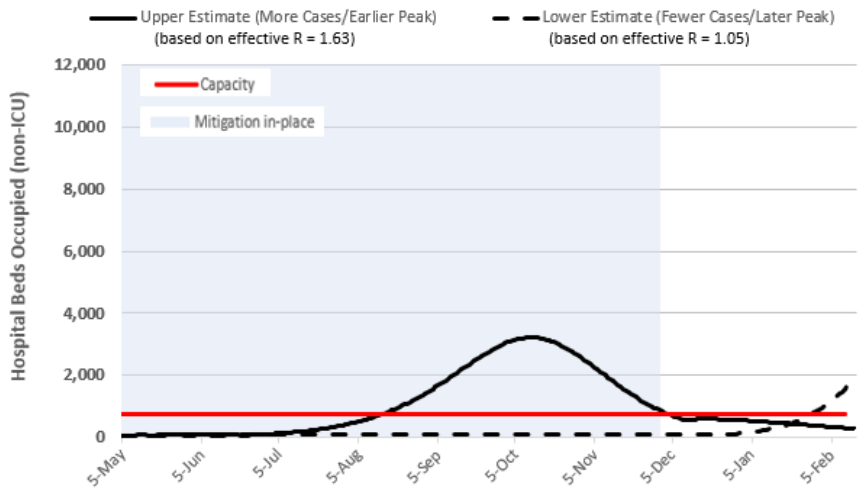
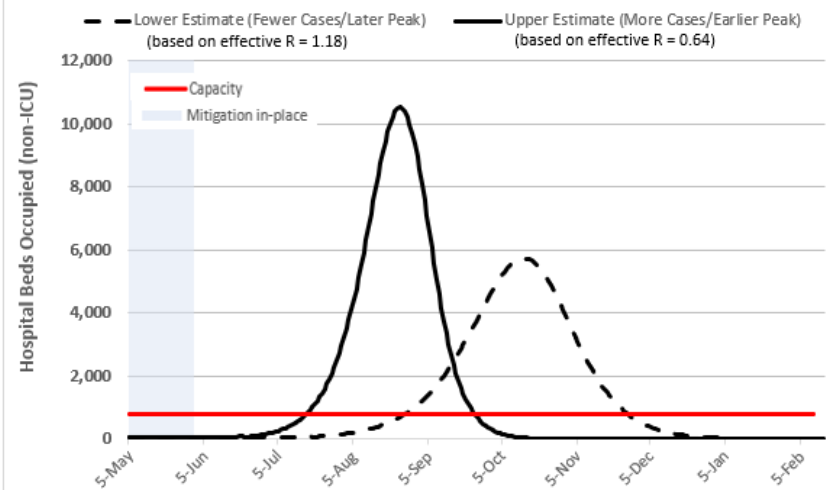

## B.2.2 Región Araucanía - ICU beds

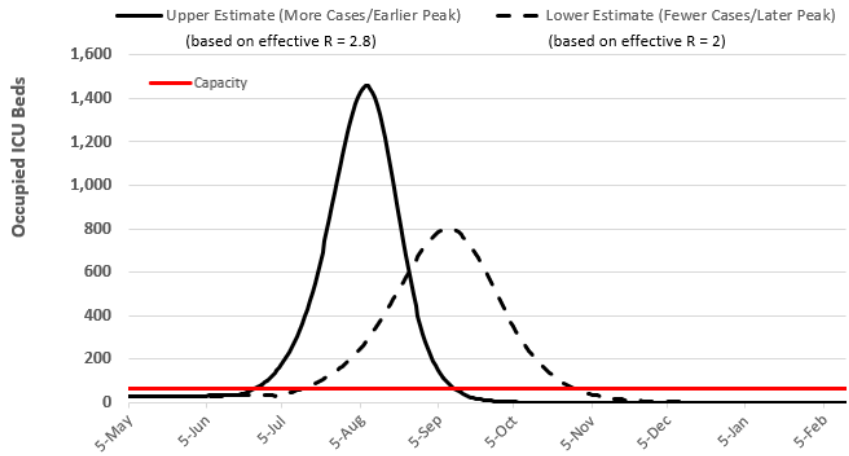
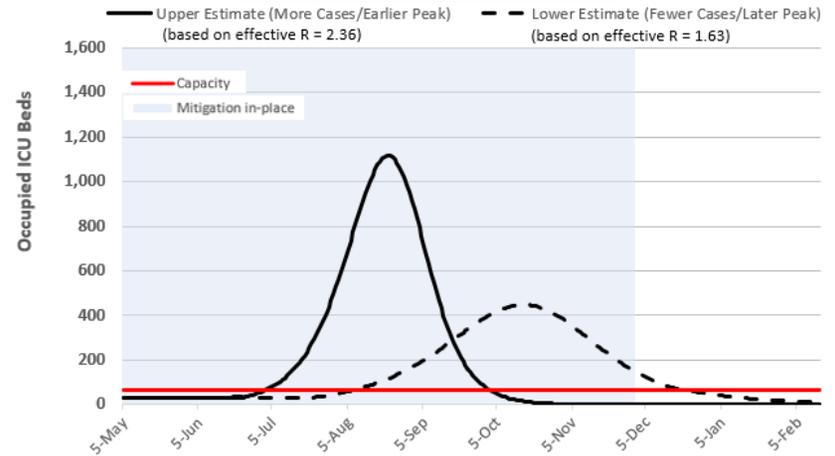
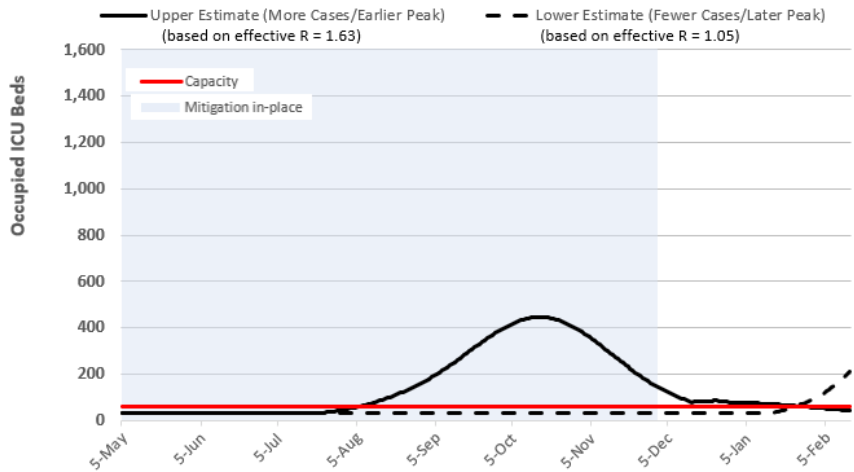
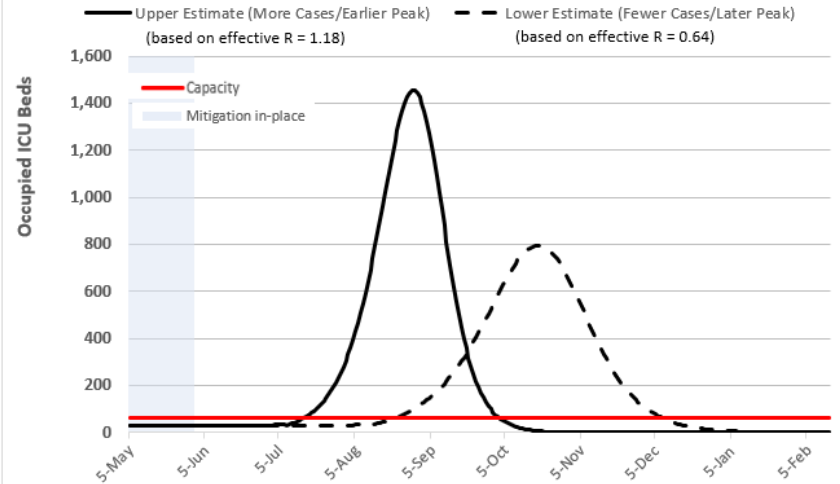

### B.2.3 Región Araucanía - Ventilators

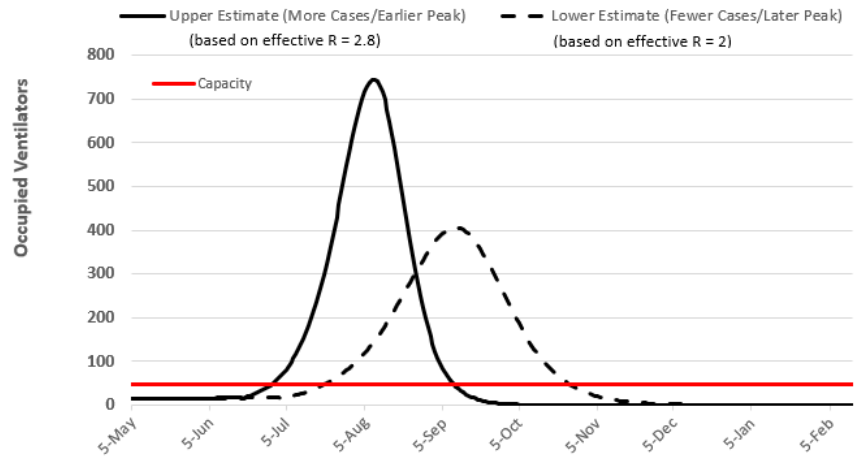
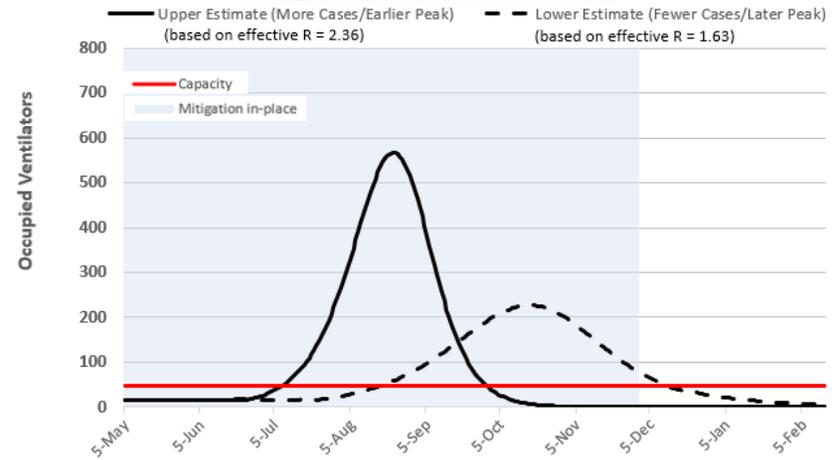
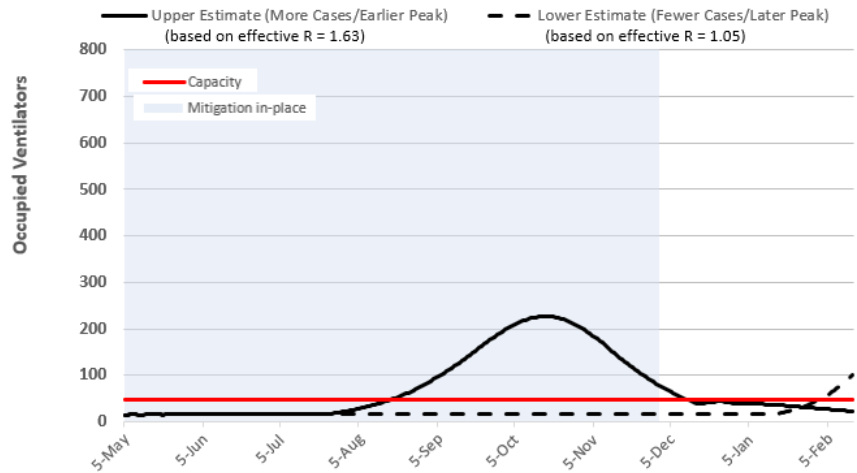
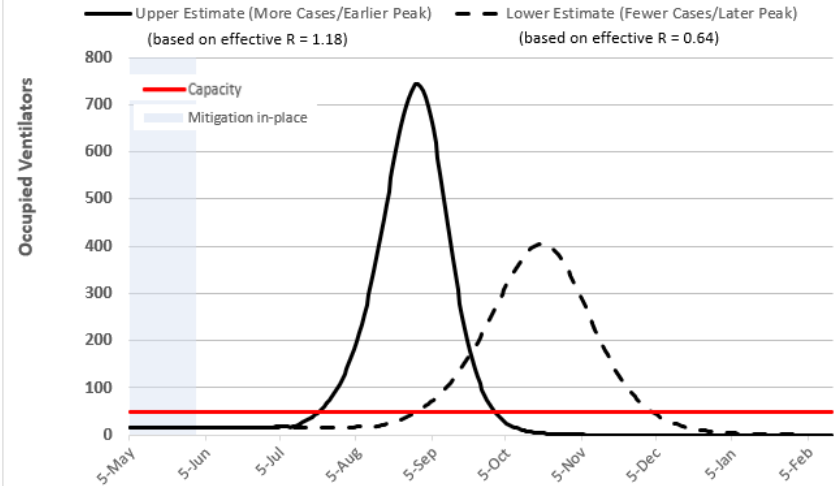

### B.3.1 Región Ñuble - Hospital beds

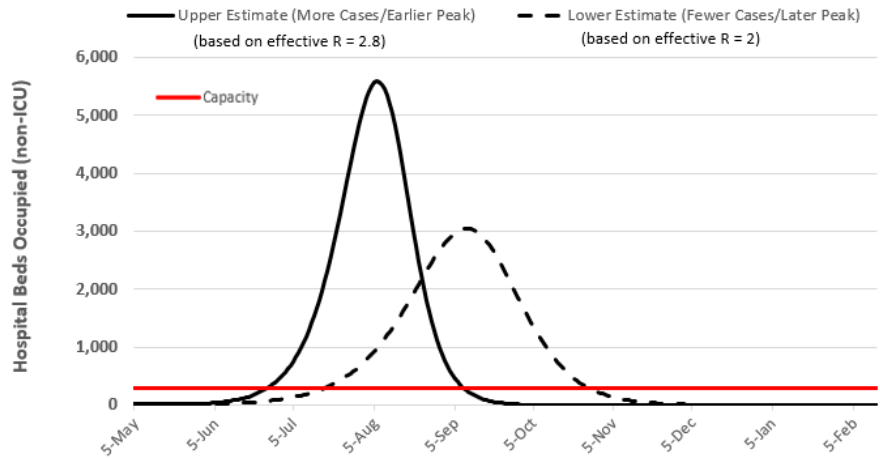

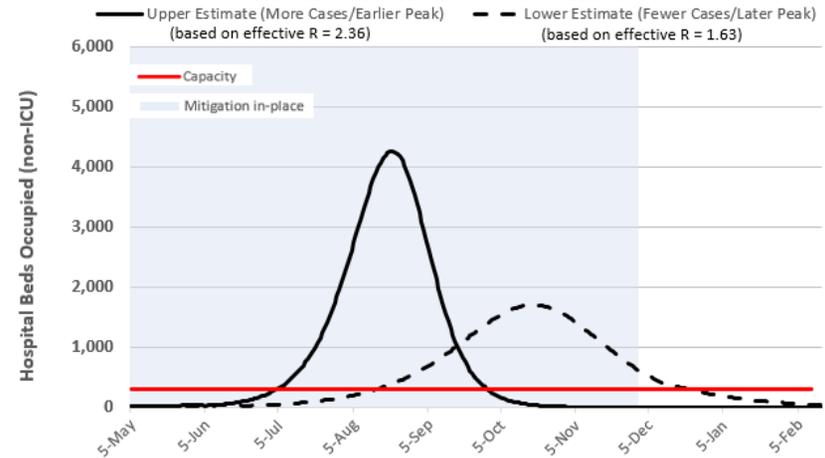

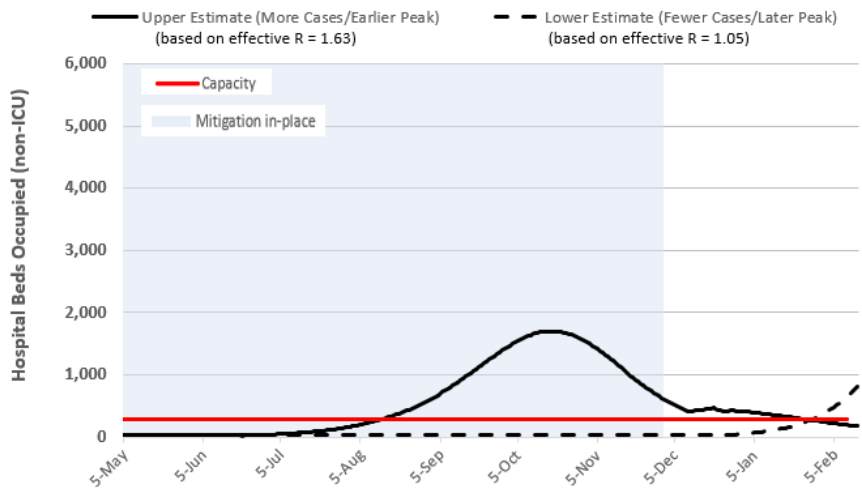

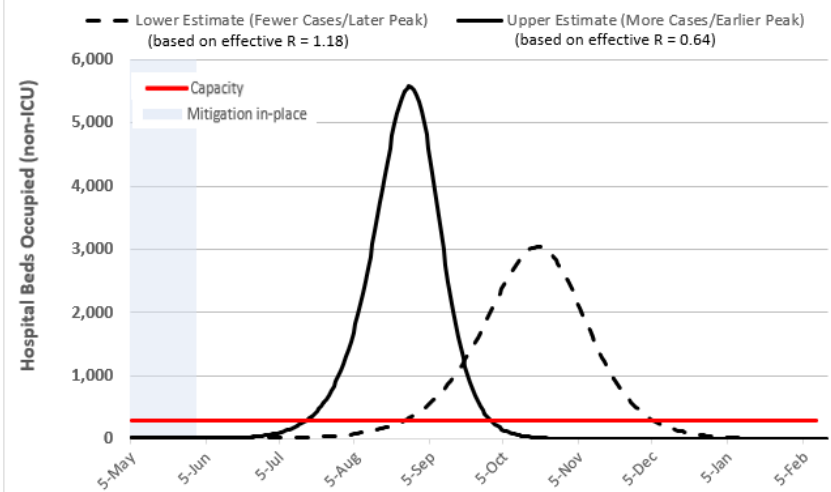

## B.3.2 Región Ñuble - ICU beds

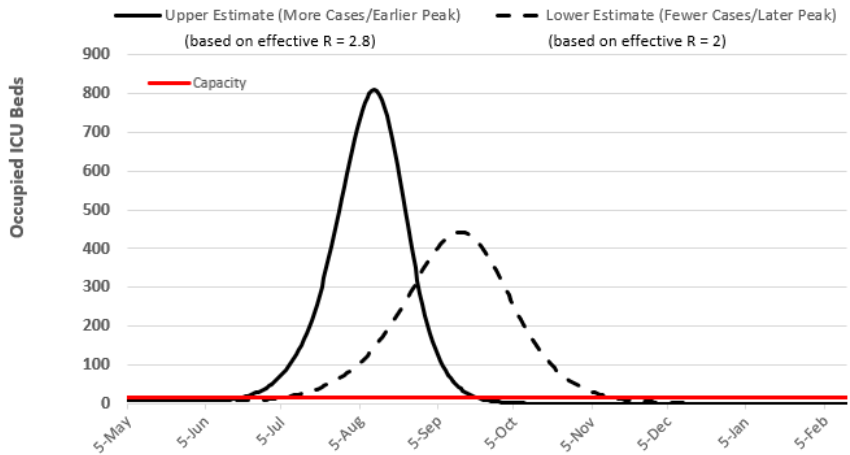
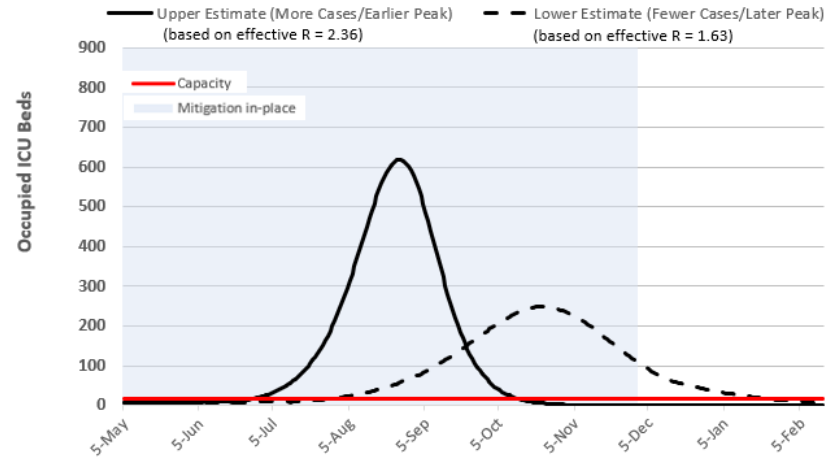
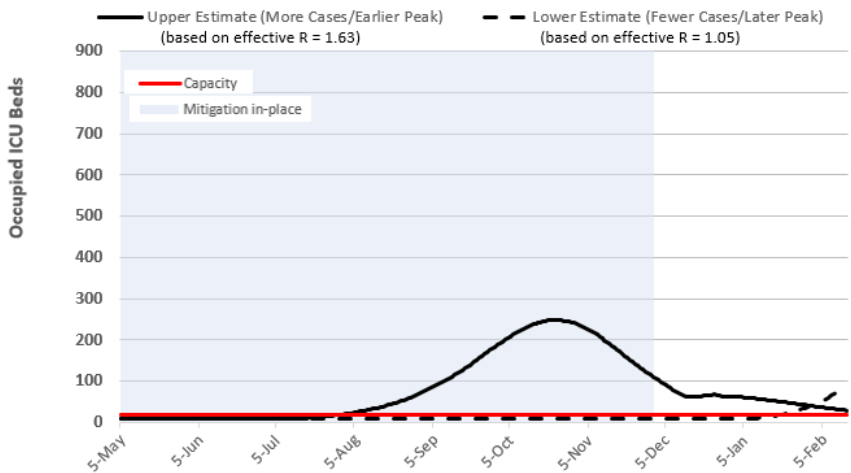
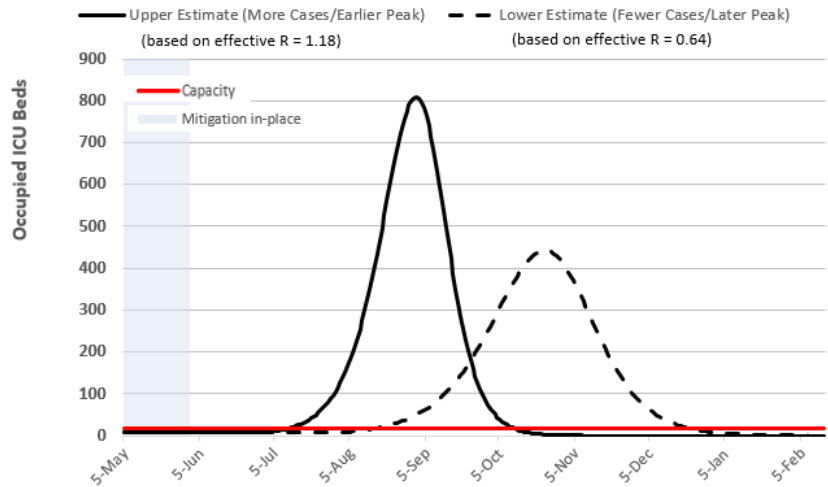

### B.3.3 Región Ñuble - Ventilators

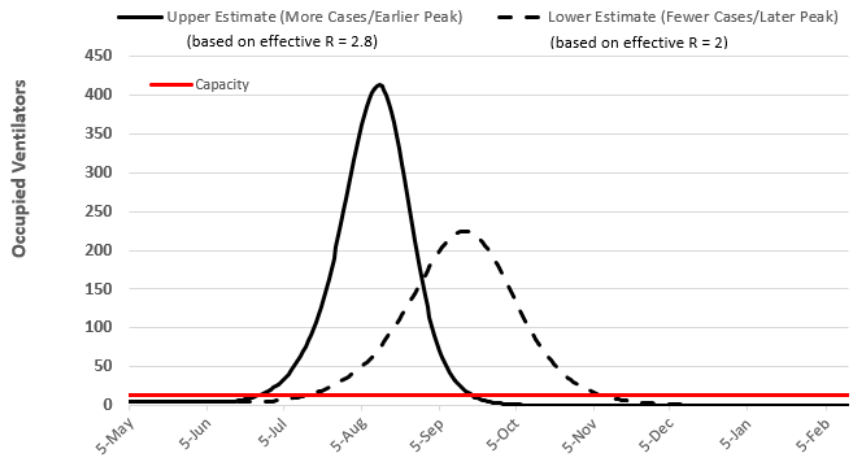
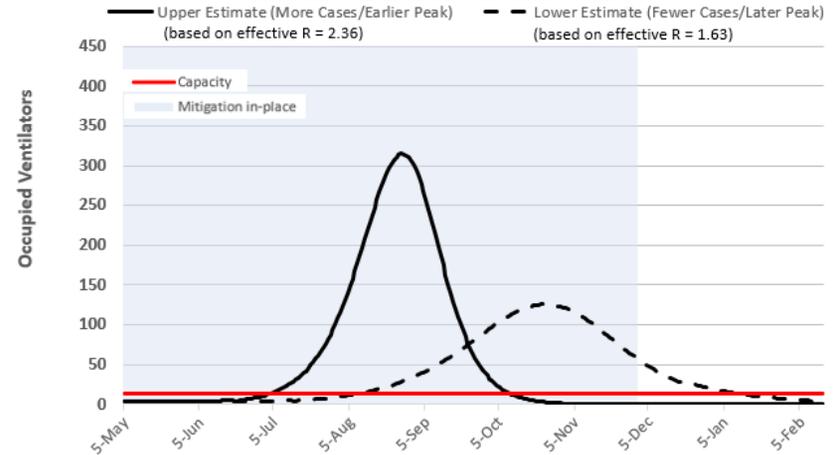
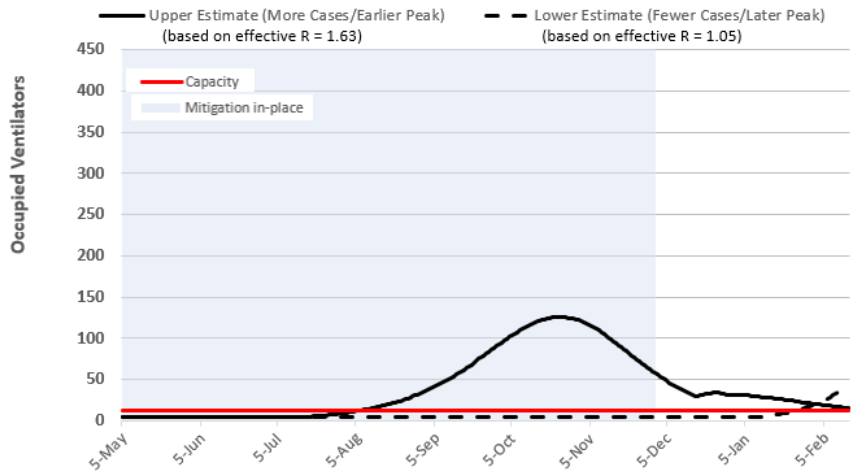
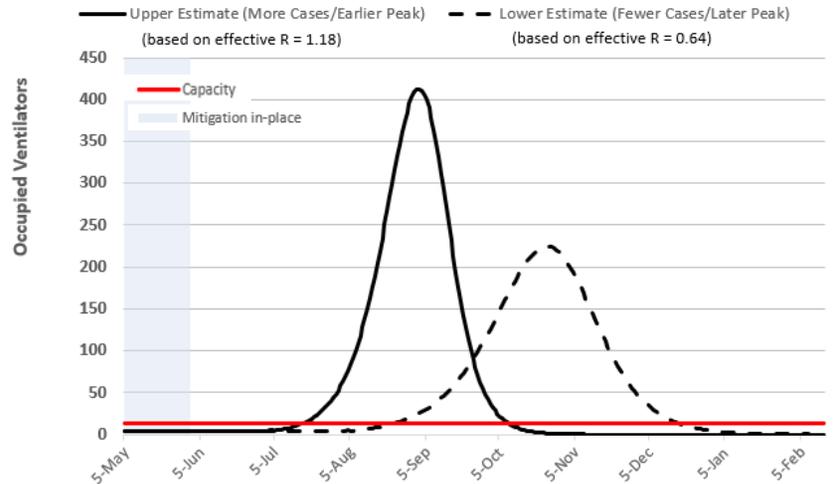

**References**

1. Ferguson N, Laydon D, Nedjati Gilani G, et al. Report 9: Impact of non-pharmaceutical interventions (NPIs) to reduce COVID19 mortality and healthcare demand. 2020.

2. Ministry of Health. Dispone medidas sanitarias que indica por brote de COVID-19. Norms 1143498, 1143591, 1746958. In: Ministry of Health Chile, editor. Santiago: Biblioteca del Congreso Nacional de Chile; 2020.

3. Ministry of Health. Dispone régimen especial de cumplimiento de jornada laboral y flexibilidad horaria por brote de coronavirus (COVID-19). Norm 1143629. 2020. https://www.leychile.cl/N?i=1143629&f=2020-03-20&p=.

4. Instituto Nacional de Estadísticas. Estimaciones y proyecciones de la población de Chile 1992-2050. 2017. https://www.censo2017.cl/ (accessed April 2 2020).

5. Ministry of Health Chile. Cifras Oficiales COVID-19. 2020. https://www.gob.cl/coronavirus/cifrasoficiales/ (accessed April 2 2020).

6. Latorre R, Sandoval G. El mapa actualizado de las camas de hospitales en Chile. La Tercera. 2020 March 31.

7. Wunsch H, Wagner J, Herlim M, Chong D, Kramer A, Halpern SD. ICU occupancy and mechanical ventilator use in the United States. *Critical care medicine* 2013; **41**(12).